 \documentstyle[preprint,aps,tighten]{revtex} 
  \newcommand{\beq}{\begin{equation}}
  \newcommand{\eeq}{\end{equation}}
  \newcommand{\beql}[1]{\begin{equation}\label{eq:#1}}
  \newcommand{\beqa}{\begin{eqnarray}}
  \newcommand{\eeqa}{\end{eqnarray}}
  \newcommand{\beqas}{\begin{eqnarray*}}
  \newcommand{\eeqas}{\end{eqnarray*}}
  \newcommand{\bra}[1]{\langle#1|}
  \newcommand{\ket}[1]{|#1\rangle}
  \newcommand{\braket}[1]{\langle#1\rangle}
  \newcommand{\ps}{\psi}
  \newcommand{\ep}{\epsilon}
  \newcommand{\eq}[1]{(\ref{eq:#1})}
  \newcommand{\mb}{\mbox}
  \newcommand{\de}{\delta}                                             %
  \newcommand{\ph}{\phi}                                               %
  \newcommand{\cQ}{{\cal Q}}
\begin{document}
\draft
\preprint{}
\title{Measurability and Computability}
\author{Masanao Ozawa}
\address{School of Informatics and Sciences,
Nagoya University, Chikusa-ku, Nagoya 4648601, Japan}
\date{\today}
\maketitle
\begin{abstract}
The conceptual relation between the measurability of quantum mechanical
observables and the computability of numerical functions is re-examined.
A new formulation is given for the notion of measurability with finite 
precision in order to reconcile the conflict alleged by M. A. Nielsen 
[Phys.\ Rev.\ Lett.\ 79, 2915 (1997)] that the measurability of a certain 
observable contradicts the Church-Turing thesis. It is argued that any 
function computable by a quantum algorithm is a recursive function 
obeying the Church-Turing thesis, whereas any observable can be measured 
in principle.
\end{abstract}
\narrowtext

In the theory of quantum computing \cite{Deu85,EJ96}, 
the process of computation consists of the preparation of the computing 
machine, the unitary evolution, and the measurement 
so that the mathematical notion of computation is intrinsically 
related to the notion of measurement in quantum mechanics.
Since the Church-Turing thesis \cite{Chu36,Tur36} defines the notion of 
computable functions independent of the notion of measurement 
of quantum mechanical observables,
it is an interesting problem to examine whether or not those two notions 
lead to a conflict.
Recently, Nielsen \cite{Nie97} argued that the measurability of a certain 
quantum mechanical observable, called the halting observable, 
leads to the computability of a noncomputable
function and concluded that either the Church-Turing thesis 
needs revision or that only a restricted class of observables may
be realized, in principle, as measurements.
The purpose of this Letter is to re-examine Nielsen's argument and
to establish a precise conceptual relation between 
the measurability of observables and the computability of functions.

It can be pointed out that Nielsen's formulation of the measurability 
of observables does not respect sufficiently the fact that every 
measuring apparatus has only finite precision as long as it
can be constructed in a laboratory or it is finitely realizable. 
A new formulation will be given for the measurability with finite 
precision from this point of view.  
The measurability of an observable is defined naturally based 
on this concept.
Under this formulation, it is proved that the halting observable
is measurable, contrary to Nielsen's argument, without any contradiction 
with the Church-Turing thesis.
Thus, Nielsen's argument is not acceptable.
It will be also argued that any function computable by a quantum algorithm
is a computable function obeying the Church-Turing thesis from a
proof theoretical point of view due to Church \cite{Chu36}.

Using a noncomputable function $h(x)$, called the halting 
function \cite{fn:1},
Nielsen has defined the halting observable $\hat{h}$ by
\beql{1107a}
\hat{h}=\sum_{x=0}^{\infty}h(x)\ket{x}\bra{x},
\eeq
where $\{\ket{x}\mid x=0,1,\ldots\}$ is an orthonormal basis 
for the state space of some physical system.
It is assumed that all the states $\ket{x}$ may be prepared, 
in principle.

His argument runs as follows.
Suppose that it is possible to construct a measuring
device to measure the observable $\hat{h}$.
Then one performs the following procedure: 
Construct the measuring apparatus to measure $\hat{h}$,
prepare the system in the state $\ket{x}$, and
perform the measurement.
With probability one the outcome of the measurement will be $h(x)$.
This gives a procedure for computing the halting function.
If one accept the Church-Turing thesis, this is not an acceptable
conclusion, since the halting function is not computable.

Nielsen rephrases the measurability of an observable by the existence
of a realizable measurement.
However, every realizable measurement is limited with a finite 
precision and hence the correspondence between observables and 
measurements are not one-to-one; an observable can be measured 
with various precisions and a measuring apparatus measures
many observables in a given error limit.
It follows that the notion of measurability should respect correctly
the fact that every realizable measurement has a finite precision.

We define the measurability of an observable as follows \cite{fn:2}.

{\em Suppose that a measuring apparatus operates on a system
in states $\ket{\ps_{i}}$ with $i=1,\ldots,N$ where $N$ is an arbitrary
finite integer and let $a_{i}$ be the experimentally determined 
average value of the outcomes of the apparatus 
in the state $\ket{\ps_{i}}$ with the error $\ep_{i}$.
We say that this measuring apparatus measures an observable
$A$ (with its own error limit) if
\beql{724a}
|\bra{\ps_{i}}A\ket{\ps_{i}}-a_{i}|<\ep_{i}
\eeq
for all $i=1,\ldots,N$.
If such a measuring apparatus exists, the observable $A$
is said to be measurable with precision 
$\{(\ket{\ps_{i}},\ep_{i})|\, i=1,\ldots,N\}$.
If this is the case for arbitrarily given 
$\{(\ket{\ps_{i}},\ep_{i})|\, i=1,\ldots,N\}$,
the observable $A$ is said to be measurable. 
Here, we assume that all $\ket{\ps_{i}}$ are chosen from the domain of $A$. 
}

Now, according to the above definition, we have the following

{\bf Theorem: }
{\em Let $A_{1},A_{2},\ldots$ be a sequence of bounded measurable
observables.  If an observable $A$ is 
the limit of $A_{1},A_{2},\ldots$ in the sense that we have
\beql{724b}
\lim_{n\to\infty}\bra{\ps}A_{n}\ket{\ps}=\bra{\ps}A\ket{\ps}
\eeq
for all states $\ket{\ps}$ in the domain of $A$, 
then $A$ is measurable} \cite{fn:3}.

The proof runs as follows.  For all $i=1,\ldots,N$, let $\ket{\ps_{i}}$ be
an arbitrary state in the domain of $A$, let $\ep_{i}$ be an arbitrary
positive number.
From \eq{724b} there is a number $\nu$ such that 
\beql{724c}
|\braket{\ps_{i}|A_{\nu}|\ps_{i}}-\braket{\ps_{i}|A|\ps_{i}}|
<\frac{\ep_{i}}{2}
\eeq
for all $i=1,\ldots,N$.
Since $A_{\nu}$ is measurable, there exists
a measuring apparatus with the average outcomes $a_{i}$ and 
the errors $\ep_{i}/2$ in the states $\ket{\ps_{i}}$ satisfying
\beql{724d}
|\braket{\ps_{i}|A_{\nu}|\ps_{i}}-a_{i}|<\frac{\ep_{i}}{2}.
\eeq
From \eq{724c} and \eq{724d} we have
\beql{724e}
|\braket{\ps_{i}|A|\ps_{i}}-a_{i}|<\ep_{i}.
\eeq
It follows that the above measuring apparatus measures $A$ with 
precision $\{(\ket{\ps_{i}},\ep_{i})|\ i=1,\ldots,N\}$.
Since this precision is arbitrarily given, we conclude that
$A$ is measurable. 

Note that an observable on a finite dimensional state space may have a
measuring apparatus with absolute precision if it is measurable, 
but it is not the case in general. 

In order to prove the measurability of the halting observable,
we shall first note that there is no conflict between
the computability of functions essentially defined on a finite
set and the measurability of corresponding observables.

For any natural number $n$ the truncated function $h_{n}$ is defined by
\beql{1108a}
h_{n}(x)\equiv\left\{\begin{array}{ll}
h(x)&\mb{if $x<n$;}\\
0&\mb{if $x\ge n$.}
\end{array}\right.
\eeq
Every integer-valued function defined on a finite set is computable, 
and hence
the function $h_{n}$ is computable, since its computation consists of
the decision of the inequality $x<n$ and computing the function $h(x)$
restricted to the finite set $\{0,\ldots,n-1\}$.
Thus, every $h_{n}$ is computable.
Note that $\lim_{n\to\infty}h_{n}(x)=h(x)$ whereas
$h$ is not computable;
the limit of a sequence of computable functions is not necessarily 
computable.

The corresponding truncated observable $\hat{h}_{n}$ is given by
\beql{1108b}
\hat{h}_{n}=\sum_{x=0}^{n-1}h(x)\ket{x}\bra{x}.
\eeq
We can show the measurability of $\hat{h}_{n}$ as follows.
For any positive integer $n$, define an observable $\hat{N}_{n}$
by
\beq
\hat{N}_{n}=\sum_{x=0}^{n-1}x\ket{x}\bra{x}.
\eeq
Since it is assumed that every state $\ket{x}$ may be prepared,
we can take it for granted that every observable $\ket{x}\bra{x}$
can be measured with arbitrary precision; 
otherwise, we cannot know that the state $\ket{x}$ can be prepared.
It follows that finitely many observables $\ket{x}\bra{x}$ with
$x=0,\ldots,n-1$ can be measured simultaneously
  with arbitrary
precision, and hence
we can conclude that the observable $\hat{N}_{n}$ is measurable.

In order to measure $\hat{h}_{n}$, we need only to measure $\hat{N}_{n}$
first and to send the outcome to a computing machine to compute the function
$h_{n}$.  
Thus, the observable $\hat{h}_{n}$ is measurable.

In order to show the measurability of $\hat{h}$, it suffices to show
that for any state $\ket{\ps}$ in the domain of $\hat{h}$ we have
\beql{724f}
\lim_{n\to\infty}\braket{\ps|\hat{h}_{n}|\ps}
=\braket{\ps|\hat{h}|\ps}.
\eeq
By expanding $\ket{\ps}=\sum_{x=0}^{\infty}c_{x}\ket{x}$, we have
\beql{724g}
|\braket{\ps|\hat{h}|\ps}-\braket{\ps|\hat{h}_{n}|\ps}|
\le\sum_{x=n}^{\infty}|c_{x}|^{2}\to 0
\eeq
as $n\to \infty$
and hence \eq{724f} follows.
Thus, we conclude that the halting observable $\hat{h}$ is measurable.

Nielsen's argument intrinsically concludes that there is no single 
measuring apparatus which measures $\hat{h}$ 
with absolute precision under 
the Church-Turing thesis.
However, this conclusion is not surprising because almost every 
observable of infinite rank has no single measuring apparatus to 
measure it with absolute precision.

In order to prepare for this point, Nielsen also gave another argument
which takes approximate measurements into account.
His second argument runs as follows.
Suppose that it is possible to measure an observable $\hat{h}'$ 
which is close to $\hat{h}$.
Preparing the system in the state $\ket{x}$ and measuring $\hat{h}'$,
a result in the range $h(x)\pm\de$ is obtained with probability at
least $1-\ep$, for some small $\ep$ and $\de$.  
Clearly, by performing measurements of this type many times
it is possible to determine $h(x)$ with arbitrary high confidence.
Thus, approximate measurements of $\hat{h}$ give an algorithmic
means for computing $h(x)$.

In Nielsen's second argument, it is assumed that $\ep$ and $\de$
do not depend on $x$.
However, if $\de$ were dependent on $x$ and if $\de=\de(x)$ were not 
bounded in $x$, we could not conclude that the repetition of 
measurements of that type can determine $h(x)$.
In most cases, a measuring apparatus realizable in the laboratory is
a device for approximate measurement for which the measurement
accuracy is dependent on and unbounded in the input
states, so that the noise can be specified well only for the finite
number of input states as formulated by \eq{724a}.
Hence the repeated use of that device cannot determine
the true values for all input states $\ket{x}$. 
Thus, Nielsen's second argument excludes more common type of 
approximate measurements.

In general, in order to approach the precise measurement of a given
observable we need to use an infinite sequence of measuring apparatuses
with more and more stringent precisions.
Only if there is an algorithmic mean to produce the sequence of 
(mathematical models of) such measurements,
the sequence of measurements with finite precision can incorporate
with the quantum computation.
For instance, the sequence of precise measurements of $\hat{h}_{n}$ 
approaches the precise measurement of $\hat{h}$, but there is no
algorithmic mean to produce the sequence $\hat{h}_{n}$; otherwise
the halting function $h$ could be computable.
Thus, even the infinite sequence of precise measurements of $\hat{h}_{n}$ 
does not serve as an algorithmic mean for computing $h(x)$.

The difference between measurability and computability
can be illustrated more clearly by measurements of physical
constants where no essential distinction between quantum and classical 
measurements takes place.
Physical constants can be regarded, if the unit system is fixed, 
as an observable of the simplest type such that every state 
is its eigenstate.

Every rational numbers can be represented as a ratio of two
integers so that it can be measured with absolute precision
by a finitary method.
A real number is defined to be a Cauchy sequence of rational numbers 
$a_{1},a_{2},\ldots$.  
Since every $a_{n}$ can be measured with absolute precision, it follows
from Theorem that every real number is measurable.

On the other hand, a real number is computable if and only if the 
function $n\mapsto a_{n}$ from a natural number $n$ to the rational 
number $a_{n}$ is computable, where $\{a_{n}\}$ is a Cauchy sequence
of rational numbers representing $a$.
That is, every Cauchy sequence corresponds to a real number
and every computable Cauchy sequence corresponds to a computable
real number.
Some irrational numbers such as $\sqrt{2}$, $\pi$, and $e$ have
algorithms for producing sequences of irrational numbers converging 
them and hence they are computable \cite{fn:4}.

Is the value of the Planck constant in the standard unit system 
computable?
We know no algorithm to compute the Planck constant and we do not
believe in such an algorithm.
In fact, the cardinality of the set of computable numbers is countable 
infinite and has Lebesgue measure 0, so that an arbitrarily given real 
number from an interval is computable with probability 0.
Thus, the Planck constant is measurable but almost surely not computable,
since the given unit system is defined rather arbitrarily based on
the circumference of the earth, for example.

According to the Church-Turing thesis, the computability
is defined through computing machines which obey classical mechanics.
Hence, it might be thought that quantum mechanical computing can 
circumvent the Church-Turing thesis.

Deutsch\cite{Deu85} discussed this problem and concluded that 
the class of functions computable by quantum 
Turing machines is the class of recursive functions.
However, his argument appears to be too restrictive to settle the
problem.
Since his argument is based on the equivalence between deterministic
quantum Turing machines and reversible Turing machines \cite{Benn73},
his argument excludes the probabilistic quantum algorithms
such as Monte Carlo type algorithms and Las Vegas type algorithms.

In what follows, we shall argue that every quantum algorithm still does 
not contradict the Church-Turing thesis.

It is obvious that every recursive function is considered to be
computable.  Church \cite{Chu36} explained two reasons  
why the converse can be considered to be true.
The first one of them cannot apply directly to quantum computing, 
since it assumes that every algorithm should be represented by a 
sequence of expressions.  However, the second one is more general and 
runs as follows.

A function $y=f(x)$ of the natural numbers is said to be
representable in a formal system including the arithmetic, if there
is a formula $\ph(x,y)$ in that system 
such that for any natural numbers $m,n$ and the symbols $\mu,\nu$ 
in that system representing $m,n$ respectively, 
if $m=f(n)$ then the formula $\ph(\mu,\nu)$ is provable and 
if $m\not=f(n)$ then the negation of $\ph(\mu,\nu)$ is provable.
Then, it is well-known that if the system is consistent,
the class of representable functions coincides
with the class of recursive functions.
Church mentioned this fact as one of two grounds of his thesis.
Although the Church-Turing thesis does not define the notion of 
algorithm explicitly, the above fact suggests implicitly a condition
that every algorithm should satisfy.

It is plausible that any quantum algorithm $\cQ$ is associated with a 
mathematical proof to justify that $\cQ$ computes some function $f$.
Since a quantum algorithm depends on the theory of Hilbert spaces
and operators, the proof may use theorems in much wider fields 
than pure arithmetic, but they can be formalized in the 
Zermelo-Fraenkel set theory with the axiom of choice (ZFC), 
an axiomatic foundation of the current mathematics.
Then, we require that any quantum algorithm can be expressed by 
a numeralwise provable formula in the following sense:
By the formalization of the quantum algorithm $\cQ$ and its proof
in ZFC we can construct a formula $\ph(x,y)$ in ZFC such that 
if $m=f(n)$ then $\ph(\mu,\nu)$ is provable and if $m\not=f(n)$ then
the negation of $\ph(\mu,\nu)$ is provable \cite{fn:8}.
If this is the case, the function $f$ is representable by $\ph(x,y)$ 
in ZFC so that $f$ is a recursive function.
Thus, we can conclude that every function $f$ computable by a quantum 
algorithm is a recursive function.
Nielsen's procedure is proved to compute the halting function $h$,
but the procedure cannot be expressed by a numeralwise provable formula,
since $h$ is not representable unless ZFC is inconsistent.
Hence, Nielsen's procedure cannot be considered as a quantum 
algorithm \cite{fn:5}.

Currently, quantum algorithms are represented by
quantum Turing machines \cite{Beni80,Deu85,BV97}
or quantum circuit families \cite{Deu89,Yao93}. 
Since a quantum Turing machine has a tape with infinite length,
it cannot be considered as a finitely realizable physical system,
but as an infinite sequence of finitely realizable physical systems.

For a computing process using a quantum Turing machine to be
a quantum algorithm, it is necessary that the
quantum transition function, or the matrix elements of the 
time evolution unitary operator in the computational basis,
has values in the computable complex numbers.
In this case, every function computable by a quantum
Turing machine is a recursive function.

On the other hand, quantum circuits have only a finite dimensional 
state space, and  hence a quantum algorithm to compute a numerical
function with arbitrarily large inputs is represented by a family
$\{\cQ_{n}\}$ of quantum circuits of the arbitrary input length $n$.
The computing performed by a quantum circuit family is a quantum algorithm
if and only if the matrix elements 
$\braket{x|U_{n}|x'}$ in the computational basis of the unitary 
operator $U_{n}$ representing each quantum circuit 
$\cQ_{n}$ are computable complex numbers computable in $x,x',n$.
Such a quantum circuit family is called a uniform quantum circuit 
family \cite{fn:6}.
In this case, it can be concluded that every function computable 
by a uniform quantum circuit family is a recursive function.
Shor's algorithms for discrete logarithm and factorization are
considered to be represented by uniform quantum circuit families
\cite{Sho97}.

Computer science treats two levels of computability: the computability
defined by the Church-Turing thesis and the efficient computability
formalized in the computational complexity theory.  Accordingly,
the physics of computing has the following four levels of reality 
of the physical entities.  

(1) {\em Experimental reality} represented by finitely realizable objects
such as rational numbers, finite dimensional rational matrices, and
measuring devices.

(2) {\em Efficiently computable reality} represented by a polynomial time
computable sequence of finitely realizable objects such as polynomial
time computable real numbers, discrete Fourier transforms, and 
Shor's algorithm.

(3) {\em Computational reality} represented by a constructive sequence of
finitely realizable objects such as computable real numbers,  
quantum Turing machines, uniform quantum circuit families.

(4) {\em Mathematical reality} represented by a Cauchy sequence of 
finitely realizable objects such as real numbers, observables, and states.

Mathematics represents infinity by a finite length formula with a finitarily
decidable grammar, because of the limitation of our ability of reading
the formula.  Physics has the same footing as mathematics in the
experimental level by the limitation of our ability of reading the 
scale of the meter.  Thus, physics also represents infinity by a finite
length mathematical formula.  
According to this constraint, any computation of an infinite function 
and any observable of infinite rank have no experimental reality 
and need a help from the rules of the language.
This justifies our requirement that a quantum algorithm can
be expressed by a numeralwise provable formula in a formal language.
From this, we have shown that any function computable by a quantum
algorithm is a recursive function, whereas any observable is
measurable \cite{fn:7}.

I thank Horace Yuen for stimulating discussions on this paper.



\begin{references}
\bibitem{fn:1} {Although $h$ is specified to be the halting function 
tentatively, the reasoning which follows applies to any noncomputable
functions.}
\bibitem{fn:3} {For mathematically complete treatment, note that the 
analogous assertion holds also for the limits of nets of observables.}
\bibitem{fn:8} From a formal point of view, the quantum algorithm $\cQ$
can be identified with the formula $\ph(x,y)$ and the status of
$\cQ$ can be regarded as an intuitive expression of $\ph(x,y)$.
\bibitem{fn:5} {In fact, his proof only shows that his procedure
has the output $h(n)$ for the input $n$.  Since the function $h$ is defined
nonconstructively, we have no way to produce the formula $\ph(x,y)$
to represent $h$.}
\bibitem{fn:6} {By a uniform family, it is often meant that the
code of $\cQ_{n}$ is polynomially computable, but we require here
simply that it be computable.}
\bibitem{fn:7} {This follows from, for instance, the fact that the set 
of computable matrices is dense in the full operator algebra with the 
weak operator topology.}
\end{references}
\end{document}